\documentclass[a4paper]{jpconf}
\usepackage{graphicx}
\usepackage{amsfonts,amsmath,amssymb,bm,graphicx,cite,makeidx,multicol,color}
\usepackage{slashed}
\begin{document}
\title{Neutrino evolution and quantum decoherence}

\author{Konstantin Stankevich}

\address{Department of Theoretical Physics, Moscow State University, 119992 Moscow, Russia}

\ead{kl.stankevich@physics.msu.ru}

\author{Alexander Studenikin}

\address{Department of Theoretical Physics, Faculty of Physics,	Lomonosov Moscow State University, Moscow 119991, Russia}

\address{Joint Institute for Nuclear Research, Dubna 141980, Moscow Region, Russia}

\ead{studenik@srd.sinp.msu.ru}

\begin{abstract}

 Neutrino interactions with an external environment can influence the neutrino oscillation pattern and the oscillations can be damped  as a result of the neutrino quantum decoherence. In particular, the quantum decoherence of neutrino states engendered by the neutrino radiative decay accounting for the nonstandard interactions (NSI) leads to the suppression of flavor neutrino oscillations in the solar neutrino fluxes.
 \end{abstract}

The neutrino oscillation patterns can be modified by neutrino interactions with external environments including electromagnetic fields that can influence on neutrinos in the case neutrinos have nonzero electromagnetic properties \cite{Giunti_Studenik}. The phenomenon of neutrino oscillations can proceed only in the case of the coherent superposition of neutrino mass states. An external environment can modify a neutrino evolution in a way that conditions for the coherent superposition of neutrino mass states are violated. \cite{Stankevich_Studenikin,Stankevich_Studenikin_2,Stankevich_Studenikin_3}.
 Within reasonable amount of the performed studies the method based on the Lindblad master equation \cite{Lindblad, Gorini_Kossakowski} for describing neutrino evolution has been used. This approach is usually considered as the most general one that gives a possibility to study neutrino quantum  decoherence as a consequence of interactions of a neutrino system with matter, nonstandard interactions, quantum foam and quantum gravity (see \cite{Joao_Coelho2} and the corresponding references).

In the approach that uses the Lindblad master equation the quantum decoherence is described by the free dissipative parameters that can be constrained (or found) by the  experimental data on neutrino fluxes.  On the contrary, these parameters are fixed {\it ad hoc}. Currently the  long-baseline neutrino experiments provide constraints on the   decoherence parameter on the order of $\Gamma_1 \sim 10^{-24}$ GeV \cite{Coloma}.

In  \cite{Stankevich_Studenikin,Stankevich_Studenikin_2,Stankevich_Studenikin_3} we have proposes and developed a new theoretical framework, based on the the quantum field theory of open systems \cite{Breuer_Pettrucione}, for the neutrino evolution in an external environment. We implemented the proposed approach to the consideration of a new mechanism of the neutrino quantum decoherence that appears due to neutrino radiative decay in the thermal background of electrons and photons. Within our approach we have obtained the explicit expressions of the dissipative parameters as functions of the characteristics of an external environment and also on the neutrino energy. We also apply the developed approach to the study of the influence of dark photons on the neutrino quantum decoherence.

In this short note we consider a new possibility to constrain  the neutrino radiative decay  rate. In the standard model, the neutrino radiative decay is suppressed by the GIM mechanism (see, for instance, \cite{Giunti_Studenik}), but the rate can be increased by  nonstandard interactions.
The main problem of the direct detection of the radiative decay is due to high luminosity of the astrophysical backgrounds that is usually much greater than the neutrino luminosity \cite{Oberauer:1993yr,Birnbaum:1997ji,Mastrototaro:2019vug}. To avoid this, we propose to constrain the neutrino radiative decay by searching for neutrino quantum decoherence in the fluxes from astrophysical sources, in particular in the solar neutrino fluxes.

The detailed calculations of the influence of the radiative decay on neutrino evolution can be found in \cite{Stankevich_Studenikin,Stankevich_Studenikin_2,Stankevich_Studenikin_3}.Here we indicate only the main points. We start with the quantum Liouville equation for the density matrix of a system composed of neutrinos and an electromagnetic field

\begin{equation}
\frac{\partial}{\partial t} \rho =  - i \int d^3 x\left[ H(x),\rho \right]
\label{rho}
,
\end{equation}
where $H(x) = H_\nu(x) + H_{int}(x)$ is the Hamiltonian density of the system in the interaction picture. $H_{\tilde{\nu}} = diag(\tilde{E_1},\tilde{E_2})$ is the Hamiltonian density of the neutrino system in effective mass basis and $H_{int} (x) = j_\alpha (x) A^\alpha (x)$ describes interaction between neutrino and the electromagnetic field. $A_\alpha$ is the electromagnetic field and $j_\alpha (x)$ is the current density of neutrino

\begin{equation}
j_\alpha (x) = g \overline{\nu}_i(x)\Gamma_\alpha \nu_j(x)
\label{current0}
,
\end{equation}
where $\nu_i(x)$ is the neutrino field with mass $m_i$, $\Gamma_\alpha$ stands for vector or axial vector interaction, $g$ is the effective coupling. $\Gamma_\alpha$ and $g$ depends on specific model. Here we consider axial vector interaction $\Gamma_\alpha = \gamma_\mu (1-\gamma_5)$ as an example. Using the analogous calculations to those performed in \cite{Fabbricatore_Grigoriev,Pustoshn} the current can be expressed as

\begin{equation}
j_3 = 2 g
\left(
\begin{matrix}
0 & 1 \\
1 & 0
\end{matrix}
\right)
,
\label{current}
\end{equation}

The equation (\ref{rho}) can be formally solved (integrated). Since we are not interested in the evolution of the electromagnetic field, its degrees of freedom should be traced out

\begin{equation}
\rho_\nu(t_f) = tr_f \left( Texp\left[ \int_{t_i}^{t_f} d^4x  \left[ H(x),\rho(t_i) \right]\right] \right)
\label{w1}
,
\end{equation}
where $\rho_\nu(t) = tr_f \rho(t)$ is the density matrix which describes the evolution of the neutrino system. From (\ref{w1}) we can find the following equation for the neutrino evolution allowing for radiative decay (see details in \cite{Stankevich_Studenikin,Stankevich_Studenikin_2,Stankevich_Studenikin_3})

\begin{multline}
\frac{\partial}{\partial t} \rho_{\tilde{\nu}} (t) = - i \left[ H_{\tilde{\nu}},\rho_{\tilde{\nu}}(t) \right]
+\kappa_1 \left( \sigma_-\rho_{\tilde{\nu}} (t) \sigma_+ - \frac{1}{2} \sigma_+\sigma_-\rho_{\tilde{\nu}}(t) - \frac{1}{2} \rho_{\tilde{\nu}}(t) \sigma_+\sigma_-     \right)+\\
+\kappa_2 \left( \sigma_+\rho_{\tilde{\nu}} (t) \sigma_- - \frac{1}{2} \sigma_-\sigma_+\rho_{\tilde{\nu}}(t) - \frac{1}{2} \rho_{\tilde{\nu}}(t) \sigma_-\sigma_+ \right)
,
\label{Optic2}
\end{multline}
where

\begin{equation}
\kappa_1 = \dfrac{\Delta_{ij}}{\pi^2} g^2 (f(2\Delta_{ij})+1 )
\label{parameter1}
, \ \ \ \ \ \
\kappa_2 = \dfrac{\Delta_{ij}}{\pi^2} g^2 f(2\Delta_{ij})
\end{equation}
are the parameters that describe decoherence of the neutrino system. In astrophysical environment $f(2\Delta)\gg 1$, thus one can use $\kappa_1 \approx \kappa_2 = \kappa$. In equations (\ref{Optic2}) and (\ref{parameter1}) we used the following notations: $f(E) = \dfrac{1}{e^{E/kT_\gamma}-1}$, where $T_\gamma$ is the temperature of the external photons, $
\sigma_+ =
\left(
\begin{matrix}
0 & 1 \\
0 & 0
\end{matrix}
\right),
\sigma_- =
\left(
\begin{matrix}
0 & 0 \\
1 & 0
\end{matrix}
\right)$ and $\Delta_{ij} = \dfrac{\sqrt{(\Delta m_{ij} \cos 2\theta_{ij} - A)^2 +\Delta m^2_{ij} \sin^2 2 \theta_{ij}}}{2 E}$ is the effective the squared mass difference. Here we would like to write the rate of the radiative decay with neutrino current (\ref{current0}) in external radiation field

\begin{equation}
\Gamma_{\nu_i \to \nu_f + \gamma} = \dfrac {1 }{\sqrt{2}} \Delta_{ij} g^2 f(2\Delta_{ij})
.
\end{equation}

The solution of equation (\ref{Optic2}) is given by

\begin{equation}
\rho_{\tilde{\nu}} = \frac 12
\left(
\begin{matrix}
1 + \cos 2 \tilde{\theta}_{ij} e^{-\kappa t} & \sin^2\tilde{2\theta}_{ij}  e^{i 2\Delta_{ij} t} e^{-\kappa t/2} \\
\sin^2\tilde{2\theta}_{ij} e^{-i 2\Delta_{ij} t} e^{-\kappa t/2} &  1 - \cos 2 \tilde{\theta}_{ij} e^{-\kappa t}
\end{matrix}
\right)
\label{solution}
.
\end{equation}

From (\ref{solution}) it is easy to find the probability of the neutrino flavour oscillations

	\begin{equation}
	P_{\nu_e \to \nu_x} = \sin^2 2 \tilde{\theta}_{ij} \sin^2\left(\Delta_{ij} x\right) e^{-\kappa x/2} +
	\frac 1 2 \left( 1 - \sin^2 2 \tilde{\theta}_{ij} e^{-\kappa x/2} - \cos^22 \tilde{\theta}_{ij}e^{-\kappa x}\right)
	\label{Probability}
	.
	\end{equation}	
	
From equation (\ref{Probability}) one can see that the decoherence parameter $\kappa$ leads to the damping of the neutrino oscillations that can be in principle detected by the terrestrial neutrino observatories. Recently the effect of the neutrino quantum decoherence in the solar neutrino fluxes have been studied in \cite{deHolanda:2019tuf} and the limits on the decoherence parameters of order of $\tau < 10^{-19}$ eV have been obtained. From the above considerations it just straightforward  the neutrino radiative decay rate can be constrained on order of $\Gamma_{\nu_i \to \nu_f + \gamma}<10^{-4}$ s$^{-1}$.

\end{document}